\documentclass[12pt]{iopart}
\usepackage{graphicx}  
\begin{document}

\title[Scaling of anisotropic flow in the picture of quark coalescence]{Scaling of anisotropic flow in the picture of quark coalescence}

\author{Markus D Oldenburg (for the STAR Collaboration)\footnote{For the full author list and acknowledgments see Appendix ``Collaboration'' in this volume.}
}

\address{Lawrence Berkeley National Laboratory,
1 Cyclotron Road, Berkeley, CA 94720, USA}

\ead{MDOldenburg@lbl.gov}

\begin{abstract} 
Measurements of anisotropic flow at low ($p_T$\,$<$\,1.5\,GeV/$c$) and
intermediate (1.5\ $<$\ $p_T$\ $<$\ 5\,GeV/$c$) transverse momentum
from the STAR collaboration are reviewed. While at low $p_T$ an
ordering of elliptic flow strength with particle mass is observed, the
measured signals appear to follow number-of-constituent quark scaling
at intermediate $p_T$. The observations of higher harmonics support
this picture qualitatively, and are sensitive to specific model
assumptions.

\end{abstract}


%\submitto{\JPG}

%\maketitle

\section{Introduction} The azimuthal anisotropy of particles created
in ultra-relativistic heavy-ion collisions is commonly measured by
expanding the particle's azimuthal momentum distribution with respect
to the reaction plane in terms of a Fourier series \cite{volart}. The
obtained coefficients $v_n$ for increasing order $n$ characterize the
distribution in more and more detail. In non-central collisions
sufficient rescattering will drive the initial spatial anisotropy of
the system into a state where this spatial anisotropy is
diminished. During this evolution an azimuthal anisotropy in momentum
space is built up. Due to the self-quenching nature of this process
the signal of anisotropic flow is sensitive to an early stage in the
evolution of the system.  This picture holds for all different
harmonics, even though flow of different order might test different
time scales.

The second Fourier coefficient $v_2$, so-called elliptic flow, is
studied in detail at RHIC \cite{v2overview}. This is mainly due to its
large magnitude, which allows for a precise measurement of the
reaction plane. It was realized that this high resolution of the
second order reaction plane allowed for the measurement of higher
order anisotropies as well \cite{kolb}.

While the qualitative agreement of hydrodynamical model predictions
with anisotropic flow measurements below transverse momentum of about
$p_T = 1.5$\,GeV/$c$ \cite{hydroworks} is good, at higher momenta this
picture breaks down. Instead, quark coalescence models provide a very
intriguing explanation of the observed particle species dependence in
the intermediate $p_T$-region up to $p_T \approx 5$\,GeV/$c$
\cite{mass1}.

In this article the measurements of elliptic flow $v_2$ and higher
order flow by the STAR experiment at RHIC will be reviewed. The data
presented come from Au+Au collisions at $\sqrt{s_{NN}} =
200$\,GeV. About two millions events in the STAR main time projection
chamber (TPC) with a pseudorapidity coverage of $-1.2 < \eta < 1.2$
were analyzed.

\section[Mass ordering of elliptic flow $v_2$ at low $p_T$]{Mass ordering of elliptic flow \boldmath$v_2$ \unboldmath at low \boldmath$p_T$\unboldmath}

\begin{figure}[ht]
\center{
\resizebox{
0.9\textwidth}{!}{
\includegraphics{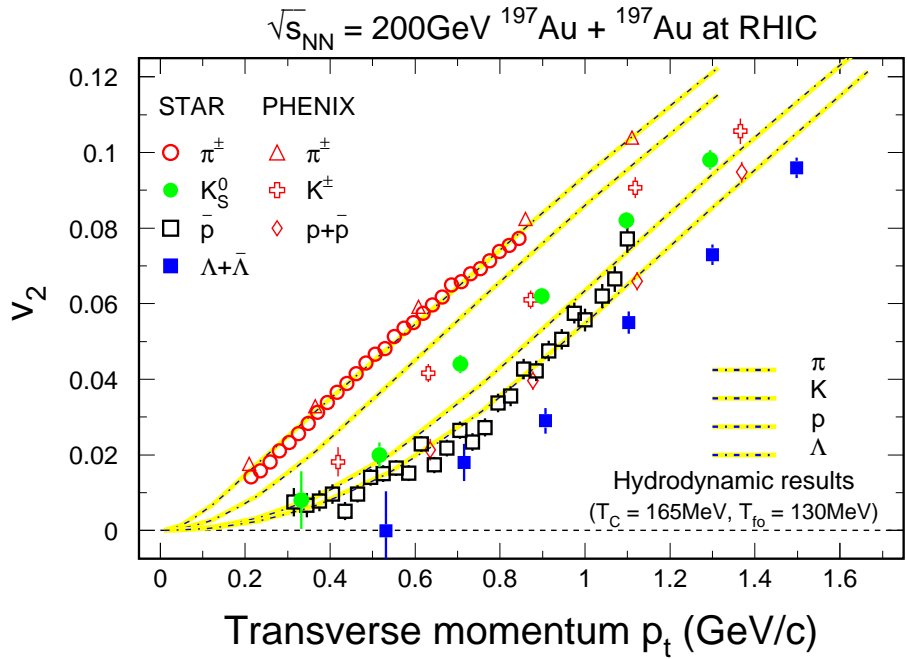}}
\caption{\label{v2pt_mass_ordering}Elliptic flow $v_2$ for different
  particle species measured by STAR and PHENIX \cite{mass2,mass1}
  compared to hydrodynamic model predictions \cite{huopriv}. The data
  indicates the expected mass ordering in this low $p_T$ region.} }
\end{figure}

In the low-$p_T$ region up to $p_T \approx 1.5$\,GeV/$c$ elliptic flow
shows an almost linear increase with transverse momentum. This feature
is present for all measured particle species, as can be seen in
Fig.~\ref{v2pt_mass_ordering}: charged pions, charged and neutral
kaons, protons and lambdas and their anti-particles -- all follow this
general behavior. A clear mass ordering effect is visible: particles
with a higher mass show a smaller $v_2$ at a given $p_T$ than
particles with lower mass.

Hydrodynamical model calculations \cite{huo,huopriv} can successfully
describe these observations in this $p_T$-region to a level of 20 --
30\,\%, attributing the mass ordering to an underlying common
transverse velocity field. These models assume a local thermal
equilibrium of the particle source. Since the mechanism of generating
anisotropic flow is self-quenching, the development of local thermal
equilibrium at an early stage is supported. Nevertheless,
since these calculations can't predict the full shape of the event,
neither the observed anisotropies at forward/backward rapidities nor
other harmonics than $v_2$, the conclusion of thermal equilibrium is
not at all solid.

It is important to note that the aforementioned hydrodynamical models
fail to describe the mass and momentum dependence of elliptic flow as
long as they don't invoke a phase transition from partonic to hadronic
degrees of freedom \cite{eos}. This sensitivity to the equation of
state justifies the importance of elliptic flow for the understanding
of hot and dense nuclear matter, including the possible creation of a
quark-gluon plasma (QGP).

\section[Meson-baryon ordering at intermediate $p_T$]{Meson-baryon ordering at intermediate \boldmath$p_T$\unboldmath}
\begin{figure}[ht]
\center{
\resizebox{
0.8\textwidth}{!}{
\includegraphics{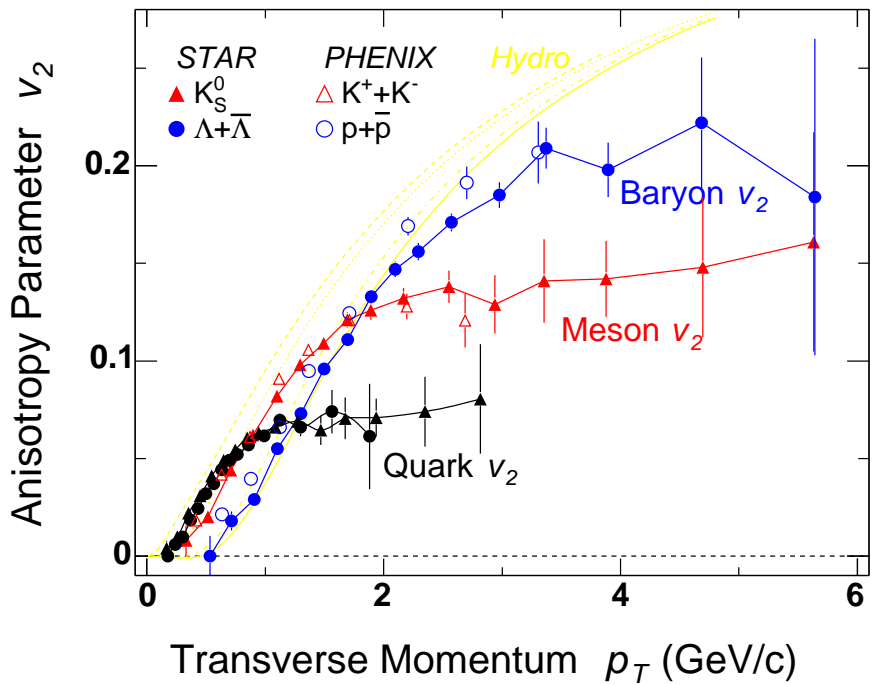}}
\caption{\label{idv2}Identified particle $v_2$, hydrodynamic
  predictions, and elliptic flow of quarks derived from
  number-of-constituent quark scaling. Figure taken from
  \cite{figpiklv2}.}  }
\end{figure}

At transverse momentum 1.5\ $<$\ $p_T$\ $<$\ 5\,GeV/$c$ the elliptic flow of
different particle species starts to deviate significantly from the
mass ordering schematics discussed above, see
Fig.~\ref{idv2}. Interestingly, different particle species seem to
exhibit a different value of saturation for $v_2$, while they fall on
top of each other if grouped into mesons and baryons. While at low
$p_T$ the lighter mesons have a larger $v_2$ for a given $p_T$, at
intermediate $p_T$ the elliptic flow of mesons is smaller than
that of baryons.

Quark coalescence models provide an elegant explanation for this
observed feature. These models assume that in this $p_T$ range
particle production is dominated by coalescing quarks: Two quarks
moving with the same momentum make up a meson with twice the momentum
of the original quarks, three quarks coalesce into a baryon with three
times the quark momentum \cite{coamodels}. Therefore, by scaling the
observed $v_2$ signal and the transverse momentum by the number of
constituent quarks $n$ one obtains the underlying quark flow.

\begin{figure}[ht]
\center{
\resizebox{
0.8\textwidth}{!}{
\includegraphics{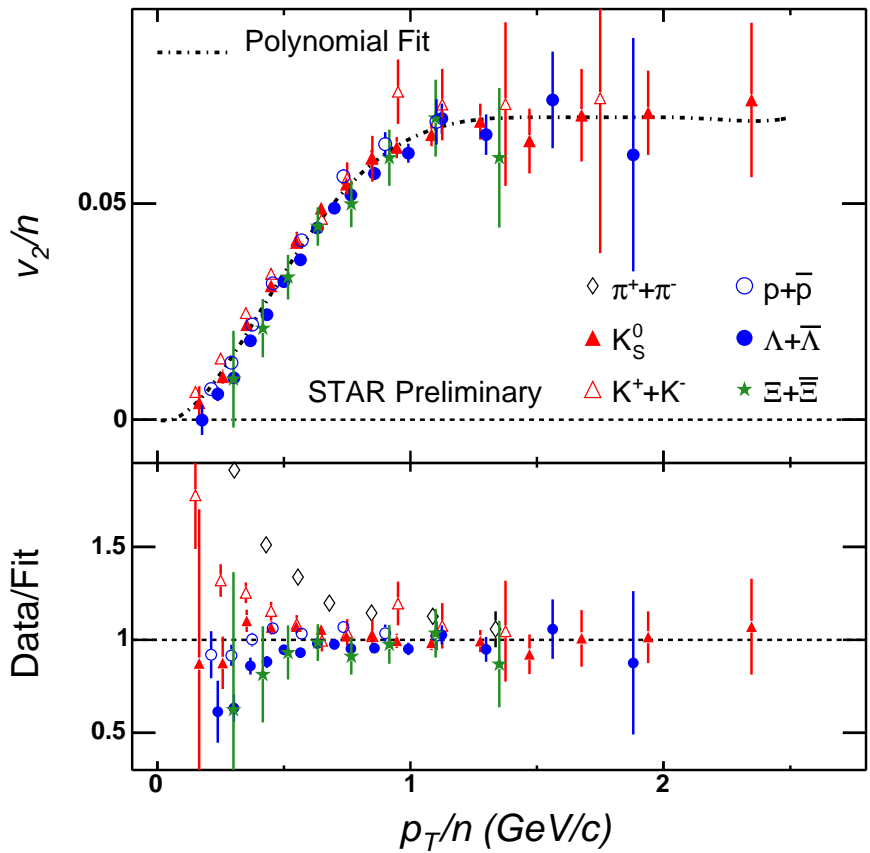}}
\caption{\label{v2_scaled}Elliptic flow for different particle species
scaled according to the number of constituent quarks of the
hadrons. The lower plot shows the ratio of the data to the
dashed-dotted fit to the data in the upper plot. Pions were not
included in this fit and are only shown in the lower panel. This
figure combines data from \cite{figv2scaling} and \cite{javier}.}}
\end{figure}

The results, as shown in Fig.~\ref{v2_scaled}, are very intriguing. In
the region $0.6 < p_T/n < 2$\,GeV/$c$ of scaled particle momentum,
elliptic flow for different particles is literally the same. Only the
pions deviate from the universal curve, which can be explained by
feed-down from resonances \cite{resonances}.

The success of these quark coalescence models in describing the
measured $v_2$ is a strong hint that the observed large anisotropic
flow builds up at the partonic stage of the system's evolution
already. On one hand this suggests that the system evolves through a
state of partonic degrees of freedom, which -- on the other hand --
suffer a lot of rescattering and therefore should be close to local
thermal equilibrium.

\section{Higher order anisotropies}
\begin{figure}[ht]
\center{
\resizebox{
0.8\textwidth}{!}{
\includegraphics{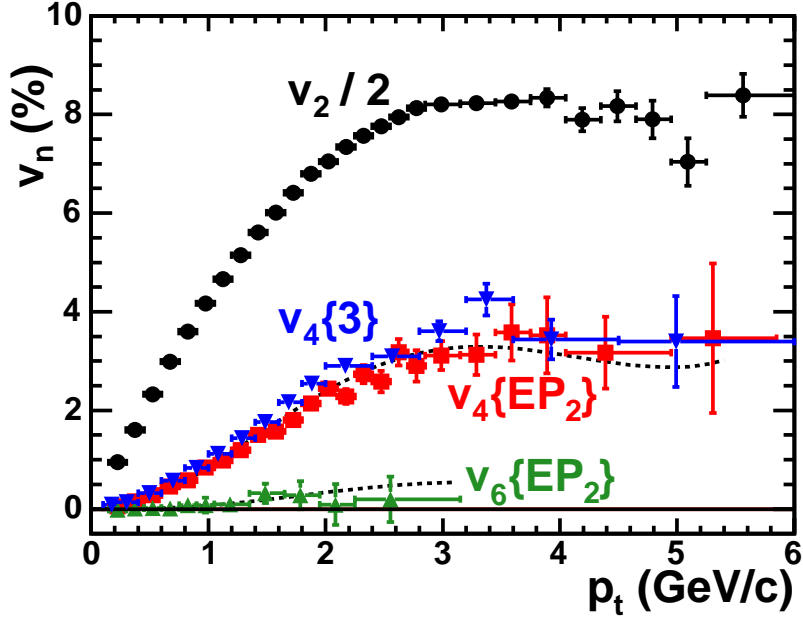}}
\caption{\label{v4}Minimum bias measurements of anisotropic flow of
charged hadrons for different harmonics. The dashed lines show
$1.2\cdot v_2^2$ and $1.2\cdot v_2^3$, respectively. Figure taken from
\cite{v1v4paper}.}
}
\end{figure}

In order to measure harmonics higher than $n=2$ we use the knowledge
of the very well determined second order reaction plane\footnote{In
this framework only harmonics which are multiples of $n=2$ -- like
$v_4$, $v_6, ..., v_{2k}, ...$ -- are accessible.} \cite{volart}. As
shown in Fig.~\ref{v4}, the coefficient for the fourth harmonic is
significantly lower than that for the second harmonic, but is non-zero
for all $p_T$. Within errors $v_6$ is equal to zero.

\begin{figure}[ht]
\center{
\resizebox{
0.8\textwidth}{!}{
\includegraphics{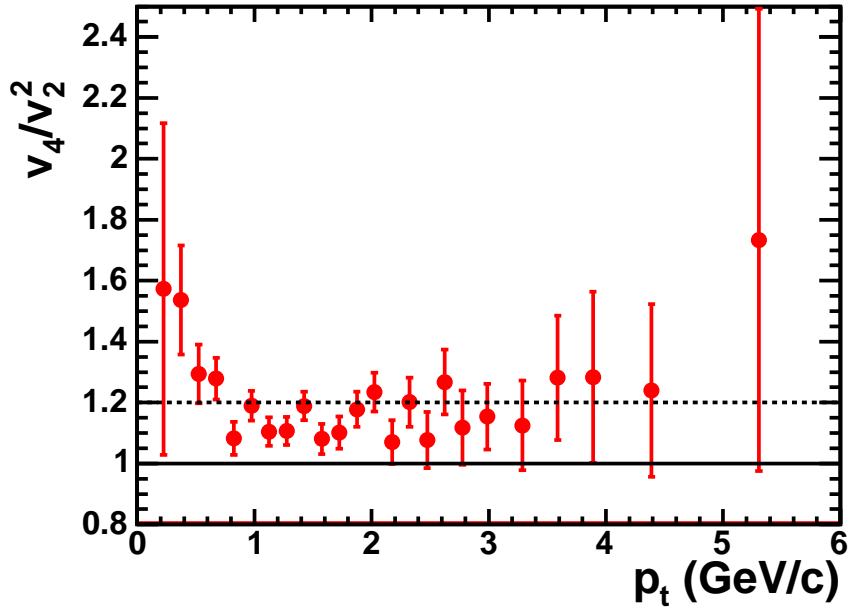}}
\caption{\label{v4v22ratio}Ratio of $v_4/v_2^2$ vs.\ $p_T$. The
  dashed straight line represents the mean of all entries at a value
  of 1.2. Figure taken from \cite{artQM}.}
}
\end{figure}

It was suggested that $v_n$ could be proportional to $v_2^{n/2}$, as
long as the $\phi$ distribution is a smooth, slowly varying function
of $\cos(2\phi)$ \cite{olli}. The ratio of $v_4$ over $v_2^2$ is shown
in Fig.~\ref{v4v22ratio} and is close to 1.2. Even though a straight
line might not be the best fit, the ratio is clearly larger than
1. According to this mean ratio the two dashed lines in Fig.~\ref{v4}
were drawn.

This very good agreement between the scaled $v_2$ and the measurements
of higher order can be also seen in the picture of parton
coalescence. Assuming a simple model \cite{kolbchen} one obtains for
the ratio $v_4/v_2^2 \approx 1/4 + 1/2\cdot(v_4^q/(v_2^q)^2)$ for
mesons and $v_4/v_2^2 \approx 1/3 + 1/3\cdot(v_4^q/(v_2^q)^2)$ for
baryons. As shown, this ratio is experimentally determined to be 1.2,
which means that the fourth-harmonic flow of quarks $v_4^q$ must be
greater than zero. One can go one step further and assume that the
observed scaling of the hadronic $v_2$ actually results from a similar
scaling occurring at the partonic level \cite{chenko}. In this case
$v_4^q = (v_2^q)^2$ and the hadronic ratio $v_4/v_2^2$ then equals
$1/4 + 1/2 = 3/4$ for mesons and $1/3+1/3=2/3$ for baryons,
respectively. Again, since this value is measured to be 1.2, even the
partonic $v_4^q$ must be greater than simple scaling and quark
coalescence models predict.
				   
\section{Summary and outlook}
Elliptic flow of identified particles was shown to exhibit a
constituent-quark-number dependence at intermediate transverse
momentum 1.5\,$<$\,$p_T$\,$<$\,5\,GeV/$c$. The quantitative measurements of
$v_2$ support model predictions for quark-number scaling which imply
hadronization by coalescing partons. Within statistical uncertainties
observations of higher order momentum anisotropies for charged hadrons
support these model predictions. The quantitative deviations of only
about 20\,\% compared to model assumptions can be easily explained in
the light of simplifications in the models while going from hadronic
to partonic scaling of $v_4$ vs.\ $v_2$.

It has to be stressed that the observed mass ordering of elliptic flow
measurements at low $p_T$ is nevertheless present and stands in no
contradiction to the described behavior at intermediate transverse
momentum.

With the large data set taken during RHIC run IV the STAR experiment
will greatly enhance the statistical significance of the presented
measurements. We will be able to test constituent-quark-number scaling
with much higher precision. It will be interesting to see the outcome
of these new measurements compared to recent developments in the
theoretical understanding of the underlying process of hadronization.

\end{document}